\documentclass[conference]{IEEEtran}

\usepackage[style=numeric,citestyle=numeric-comp,backend=biber,sorting=none,maxbibnames=99]{biblatex}
\usepackage{tikz}
\usetikzlibrary{calc,positioning,arrows,arrows.meta}
\usepackage{amsmath}
\usepackage{booktabs}
\usepackage{enumitem}
\usepackage[caption=false,font=footnotesize]{subfig}
\usepackage[hidelinks]{hyperref}

\usepackage[frozencache,cachedir=minted]{minted}
\usepackage{siunitx}
\DeclareSIUnit\sample{S}
\DeclareSIUnit[group-minimum-digits=3]\usd{USD}
\DeclareSIUnit[quantity-product = ]\percent{\char`\%}
\newcommand{\mathqty}[2][]{\qty[parse-numbers=false,#1]{#2}}

\usepackage{listings}

\newcommand*\emptycirc[1][1ex]{\tikz\draw[thick] (0,0) circle (#1);} 
\newcommand*\halfcirc[1][1ex]{%
  \begin{tikzpicture}
  \draw[fill] (0,0)-- (90:#1) arc (90:270:#1) -- cycle ;
  \draw[thick] (0,0) circle (#1);
  \end{tikzpicture}}
\newcommand*\fullcirc[1][1ex]{%
  \begin{tikzpicture}
  \draw[fill] (0,0) circle (#1) ;
  \draw[thick] (0,0) circle (#1);
  \end{tikzpicture}}

\newcommand{\dsns}[0]{\textsc{DSNS}}
\newcommand{\dsnsfull}[0]{Deep Space Network Simulator}

\usepackage{pifont}
\newcommand{\cmark}{\ding{51}}%
\newcommand{\xmark}{\ding{55}}%

\newcommand{\autobox}[1]{%
    \centering
    \resizebox{%
        \ifdim\width>\linewidth
            \linewidth
        \else
            \width
        \fi
    }{!}{#1}%
}

\begin{document}

\date{}

\title{DSNS: The Deep Space Network Simulator}

\makeatletter
\newcommand{\linebreakand}{%
  \end{@IEEEauthorhalign}
  \hfill\mbox{}\par
  \mbox{}\hfill\begin{@IEEEauthorhalign}
}
\makeatother

\author{\IEEEauthorblockN{Joshua Smailes}
\IEEEauthorblockA{University of Oxford\\
joshua.smailes@cs.ox.ac.uk}
\and
\IEEEauthorblockN{Filip Futera}
\IEEEauthorblockA{University of Oxford\\
filip.futera@cs.ox.ac.uk}
\and
\IEEEauthorblockN{Sebastian K\"{o}hler}
\IEEEauthorblockA{University of Oxford\\
sebastian.kohler@cs.ox.ac.uk}
\linebreakand
\IEEEauthorblockN{Simon Birnbach}
\IEEEauthorblockA{University of Oxford\\
simon.birnbach@cs.ox.ac.uk}
\and
\IEEEauthorblockN{Martin Strohmeier}
\IEEEauthorblockA{armasuisse Science + Technology\\
martin.strohmeier@armasuisse.ch}
\and
\IEEEauthorblockN{Ivan Martinovic}
\IEEEauthorblockA{University of Oxford\\
ivan.martinovic@cs.ox.ac.uk}}

\maketitle

\begin{abstract}

Simulation tools are commonly used in the development and testing of new protocols or new networks.
However, as satellite networks start to grow to encompass thousands of nodes, and as companies and space agencies begin to realize the interplanetary internet, existing satellite and network simulation tools have become impractical for use in this context.

We therefore present the \dsnsfull{} (\dsns{}): a new network simulator with a focus on large-scale satellite networks.
We demonstrate its improved capabilities compared to existing offerings, showcase its flexibility and extensibility through an implementation of existing protocols and the DTN simulation reference scenarios recommended by CCSDS, and evaluate its scalability, showing that it exceeds existing tools while providing better fidelity.

\dsns{} provides concrete usefulness to both standards bodies and satellite operators, enabling fast iteration on protocol development and testing of parameters under highly realistic conditions.
By removing roadblocks to research and innovation, we can accelerate the development of upcoming satellite networks and ensure that their communication is both fast and secure.

\end{abstract}

\section{Introduction}\label{sec:introduction}

As space becomes a critical component of global infrastructure, there is an increasing interest in new paradigms of communication to support the scale and complexity of upcoming networks.
Protocols are being developed to support the delay-tolerant nature of communication, and space agencies such as ESA and NASA are starting to introduce Lunar communication~\cite{israelLunaNet2020,nasaLunaNet2022,europeanspaceagencyMoonlight2024} and interplanetary networks~\cite{nasascienceMars} (cf. Figure~\ref{fig:dsns-visualizer} for a visualization of such an interplanetary network).
The presence of sporadic long-distance relay links in these networks, alongside the inherent difficulty of communication across highly distributed internet-scale networks in space, means new protocols and approaches must be taken.

Standards bodies and research organizations like the Consultative Committee for Space Data Systems (CCSDS) and Internet Research Task Force (IRTF) have been working to build standards supporting communication in the face of these new network paradigms.
For example, the Bundle Protocol (BP) provides message forwarding and delivery in networks with interrupted links~\cite{burleighBundle2022}, and the Licklider Transmission Protocol (LTP) enables reliable message transmission across individual network segments~\cite{burleighLicklider2008} -- however, it is becoming increasingly difficult to test these protocols to ensure their correct and efficient operation in the large-scale, highly distributed networks predicted to emerge in the coming decades.
Simulation tools are a crucial component of protocol development and testing, but existing offerings do not meet all the requirements for this purpose in these new networks.
There is thus a real need for new tools capable of simulating protocols at the scale of a future interplanetary internet.

\subsection{Requirements}\label{sec:requirements}

\begin{figure}
    \centering\includegraphics[width=.8\linewidth]{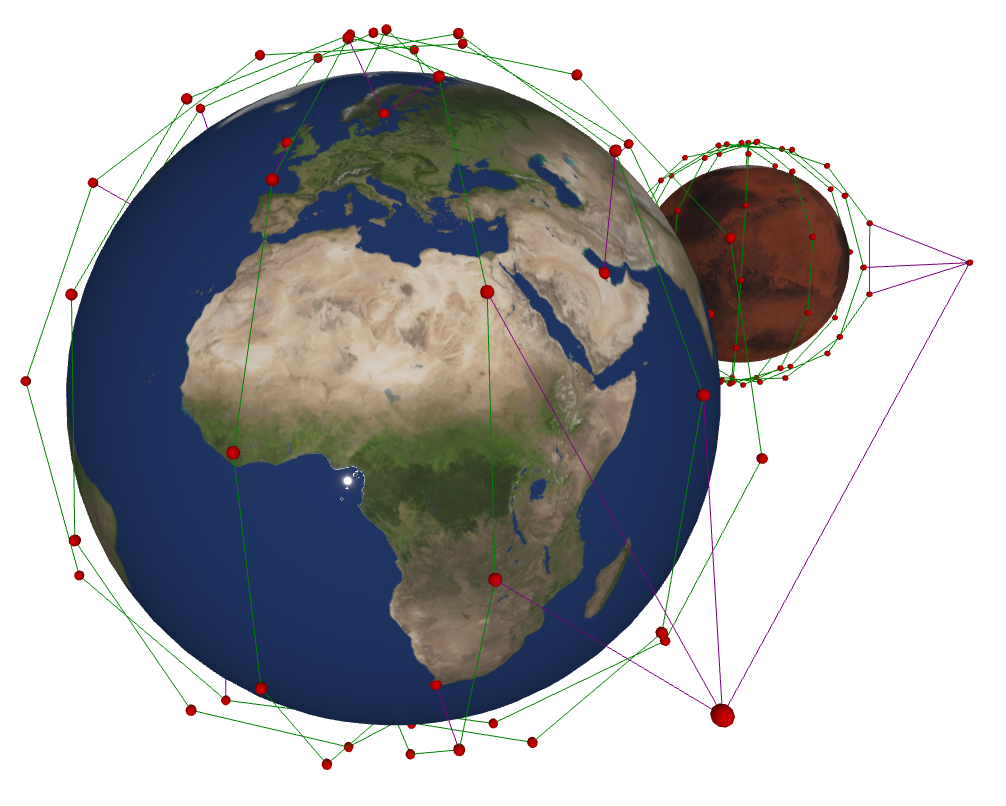}
    \caption{An interplanetary network simulated in and visualized by \dsns{}, with communication between Earth and Mars constellations via a relay link.}
    \label{fig:dsns-visualizer}
\end{figure}

\begin{table*}
    \caption{Summary of the different satellite network simulators currently available.}
    \label{tab:network-simulator-overview}
    \autobox{
    \centering
    \begin{tabular}{>{\raggedright\arraybackslash}p{5.5em}lllcc>{\raggedright\arraybackslash}p{28em}}
        \toprule
        Name                                            & Created by   & Language       & License                    & Released & Maintained & Summary \\
        \midrule
        ONE Simulator~\cite{keranenONE2009}             & TKK          & Java           & GPLv3                      & 2007     & \xmark     & Lightweight DTN network simulator. Focused on small numbers of nodes with random movement. \\[1.6em]
        NOS3~\cite{nasaOperational2025}                 & NASA         & C              & NOSA & 2019     & \cmark     & Small satellite operational simulator. Simulates flight and ground software for single missions with high fidelity. \\[1.6em]
        SpaceSecLab/ NSE2~\cite{fischerSpaceSecLab2016} & ESA          & ---            & Unreleased        & ---      & ---        & Containerized satellite simulator with integrated network simulation. Realistic and configurable, possible integration with real hardware. \\[1.6em]
            Hypatia~\cite{kassingExploring2020}             & ETH Z\"urich & C++/Python     & GPLv2/MIT                  & 2020     & \xmark     & Extension to ns-3, provides LEO constellation mobility for fixed ISLs. Now part of ``SNS3''~\cite{puttonenSatellite2015}. \\[1.6em]
        Celestial~\cite{pfandzelterCelestial2022}       & TU Berlin    & Python         & GPLv3                      & 2022     & \cmark     & LEO system testbed based on micro-VMs, supporting software emulation for LEO networks. \\[1.6em]
        StarryNet~\cite{laiStarryNet2023}               & Tsinghua     & Python         & MIT                        & 2023     & \cmark     & Simulator for integrated space and terrestrial networks, combining orbital simulation with Docker and network emulation. \\[1.6em]
        $x$eoverse~\cite{kassemxeoverse2024}            & Surrey       & Python         & Non-commercial        & 2025      & \cmark        & LEO megaconstellation simulator based on Mininet~\cite{handigolReproducible2012}. Real-time simulation of large networks including dynamic ISLs. \\[1.6em]
        Stardust~\cite{pusztaiStardust2025}             & TU Wien      & C\#            & Apache 2.0                 & 2025     & \cmark     & Scalable 3D network routing simulator, plugins to extend functionality. Fast, supports many nodes. \\[1.6em]
        SatScope~\cite{wangSatScope2025}                & NUDT China   & Python         & Non-commercial        & 2025     & ---        & LEO constellation network simulator based on VTK with a focus on satellite internet routing/coverage. \\
        \midrule
        \dsns{}                                         & Oxford       & Python         & GPLv3                      & 2025     & \cmark     & Scalable network simulator supporting arbitrary interplanetary networks. High-level protocol simulation to support development. \\
        \bottomrule
    \end{tabular}}
\end{table*}

\begin{table*}
    \caption{Comparison of the features provided by each of the satellite/DTN network simulators.}
    \label{tab:network-simulator-comparison}
    \autobox{
    \centering
    \begin{tabular}{lccccccc}
        \toprule
        Simulator                                       & Orbital Simulation (\ref{itm:leo}) & Interplanetary (\ref{itm:ipn}) & Dynamic connectivity (\ref{itm:dynamic-connectivity}) & Dynamic Timesteps (\ref{itm:dynamic-timesteps}) & Extensible (\ref{itm:extensible}) & Scalable (\ref{itm:scalable}) & Abstracted network stack (\ref{itm:abstracted}) \\
        \midrule
        ONE Simulator~\cite{keranenONE2009}            & \emptycirc & \emptycirc & \fullcirc  & \fullcirc  & \fullcirc & \halfcirc  & \fullcirc  \\
        NOS3~\cite{nasaOperational2025}                & \emptycirc & \emptycirc & \halfcirc  & \emptycirc & \halfcirc & \emptycirc & \emptycirc \\
        SpaceSecLab/NSE2~\cite{fischerSpaceSecLab2016} & \halfcirc  & \halfcirc  & \halfcirc  & \fullcirc  & \fullcirc & \halfcirc  & \emptycirc \\
        Hypatia~\cite{kassingExploring2020}            & \fullcirc  & \emptycirc & \emptycirc & \fullcirc  & \fullcirc & \halfcirc  & \halfcirc  \\
        Celestial~\cite{pfandzelterCelestial2022}      & \fullcirc  & \emptycirc & \fullcirc  & \emptycirc & \halfcirc & \halfcirc  & \emptycirc \\
        StarryNet~\cite{laiStarryNet2023}              & \fullcirc  & \emptycirc & \fullcirc  & \emptycirc & \halfcirc & \halfcirc  & \halfcirc  \\
        $x$eoverse~\cite{kassemxeoverse2024}           & \fullcirc  & \emptycirc & \halfcirc  & \emptycirc & \halfcirc & \halfcirc  & \emptycirc \\
        Stardust~\cite{pusztaiStardust2025}            & \fullcirc  & \emptycirc & \fullcirc  & \emptycirc & \fullcirc & \fullcirc  & \halfcirc  \\
        SatScope~\cite{wangSatScope2025}               & \fullcirc  & \emptycirc & \fullcirc  & \emptycirc & \halfcirc & \halfcirc  & \emptycirc \\
        \midrule
        \dsns{}                                        & \fullcirc & \fullcirc & \fullcirc & \fullcirc & \fullcirc & \fullcirc & \fullcirc \\
        \bottomrule
    \end{tabular}}
\end{table*}

We identify the following core requirements for any network simulation tool in order for it to be able to effectively aid protocol development, with a particular focus on large-scale networks and interplanetary networking:
\begin{enumerate}[label=\textbf{R\arabic*:}, ref=\textbf{R\arabic*}, leftmargin=*]
    \item\label{itm:leo} \textbf{Orbital simulation.} The simulator must be able to simulate the movement and connectivity of satellite constellations, including LEO megaconstellations.
    \item\label{itm:ipn} \textbf{Interplanetary network simulation.} The simulator must be able to handle nodes orbiting different planets, and connectivity between them.
    \item\label{itm:dynamic-connectivity} \textbf{Dynamic connectivity.} Links in the simulation must be configurable based on distance, line of sight, and/or occlusion, to ensure realistic connectivity with interruptions.
    \item\label{itm:dynamic-timesteps} \textbf{Dynamic timesteps.} In order to support long-distance interplanetary links, in addition to short-distance local links, the simulator must be able to skip forward by large timesteps when no activity is occurring, while also supporting fine-grained simulation of traffic over short-distance links.
    \item\label{itm:extensible} \textbf{Extensible.} It must be straightforward to extend the simulator to support new protocols, routing strategies, constellations, etc.
    \item\label{itm:scalable} \textbf{Scalable.} The simulator must be able to handle large numbers of nodes (at least hundreds, if not thousands or more), with large traffic volumes.
    \item\label{itm:abstracted} \textbf{Abstracted network stack.} In service of \ref{itm:extensible}, and to enable faster protocol development and testing, simulation tools benefit from supporting abstracted or reduced network stacks (potentially in addition to full-stack simulation or emulation).
\end{enumerate}

\subsection{Contributions}\label{sec:contributions}

In this paper we present the \dsnsfull{} (\dsns{}): a new network simulator optimized for interplanetary networks that satisfies all the requirements established above.
In contrast to existing offerings, explored in Section~\ref{sec:background}, \dsns{} scales well to large numbers of nodes, supports arbitrary network topologies with interplanetary links, and uses an underlying event-based simulation to enable simulations to run faster than real time.
Furthermore, \dsns{} is easily extensible thanks to its modular architecture -- new protocols can be added, removed, and swapped out, and new layers of the network stack can be implemented by simply defining new sets of rules upon which messages and events are matched.
To facilitate future research, \dsns{} has been made fully open source, released under the GNU GPLv3 license.%
\footnote{The source code and documentation can be found at \url{https://github.com/ssloxford/DSNS}.}

\section{Background}\label{sec:background}

Simulators are used in the development and evaluation of protocols in satellite and interplanetary networks, as they enable testing in networks much larger than otherwise possible, under a wide range of configurations, and without risking damage to real-world systems.
Depending on the use case, simulations may be performed purely in software, paired with hardware simulation, or connected to real-world hardware.

Several satellite network simulators are already in use; in Table~\ref{tab:network-simulator-overview} we describe each at a high level, and in Table~\ref{tab:network-simulator-comparison} we assess them against the requirements established in Section~\ref{sec:requirements}.
We justify these assessments below.
In addition to these, there are also some more generalized network simulation tools, such as ns-3~\cite{rileyns32010} and OMNeT++~\cite{vargaOMNeT2010}.
However, these do not scale well to large numbers of nodes with many connections, so we do not consider them by themselves in this paper.\footnote{There is also an extension to ns-3 called SNS3~\cite{puttonenSatellite2015} which provides satellite spot beam simulation and an implementation of the DVB-S2 protocol; however, it does not solve its scalability issues, so it is only suitable for simulating smaller networks.}

The ``ONE Simulator''~\cite{keranenONE2009} is a lightweight network simulator designed with DTNs in mind, primarily supporting scenarios with randomly moving nodes connecting based on proximity, with a well-defined extension framework.
However, it does not natively support orbital simulation or satellite movement, and scalability is limited, with tested configurations limited to approximately \num{1000}~nodes.

More recently, NASA have released their ``NOS3'' simulator, designed to be a satellite digital twin for developing and testing onboard software~\cite{nasaOperational2025}.
Although highly suited to this purpose, it is not feasible to run it at scale to test network protocols between many nodes.
ESA are also planning to release their ``Network Simulation and Emulation Environment (NSE2)'' as part of their ``SpaceSecLab''~\cite{fischerSpaceSecLab2016}, providing a Docker-based network simulation that supports highly realistic emulation of the network stack and connectivity for small numbers of nodes.
This is useful for testing implementation-specific details and small-scale mission control, but is not as practical for large satellite networks.

There has also been recent academic interest in satellite network simulators, with a range of newly released simulators since \num{2020} -- summarized in Tables~\ref{tab:network-simulator-overview} and~\ref{tab:network-simulator-comparison}.
Notably, the vast majority of these simulators either use fixed timesteps or tie the simulation to a real-world clock.
This makes the simulation of interplanetary networks particularly difficult, since they include short-distance local links alongside very long interplanetary links -- to support both of these, a very small timestep will need to be chosen, increasing the simulation runtime to impractical levels.
Event-driven simulators including ``Hypatia'', ``ONE Simulator'' and ``NSE2'' do not do this, but still struggle when large amounts of events are generated in a short timespan, triggering unnecessary position updates.
This can be improved upon by only recomputing connectivity graphs once a minimum time delta has passed.

Also notable is that many of the simulators make use of virtualization or emulation to simulate real network stacks -- this can be useful when protocols are being tested at the implementation level, but is not necessary for the vast majority of research and development tasks, slowing down development by requiring a full implementation of the protocol even at the earliest stages of testing.
Furthermore, interplanetary networks often involve new protocols across the whole stack which do not always have mature implementations, so it may not even be possible to achieve a full network stack, especially as many components will differ from the traditional IP stack.
This is of particular concern to simulators like ``$x$eoverse'' and ``Celestial'', as their underlying virtualization platforms make use of the host network stack -- this is unlikely to match the tested networks.
Similarly, ``Stardust'' cannot be used for traffic simulation or emulation, since it focuses instead on deploying computation tasks for edge computing cases.

Finally, we highlight that none of the existing simulators support interplanetary networks and many of them only have limited scalability.
Both of these features are critical to ensure communication protocols and management systems can handle the high latencies and frequent interruptions involved in interplanetary settings, and adapt to the predicted scale of these networks in coming decades.

Although these simulators are highly capable for the tasks they were designed for, none provide all the features required for effective interplanetary network simulation.
Some struggle to scale to large numbers of nodes, some do not simulate mobility, and others do not support dynamic links or multi-planet systems, severely limiting the range of network configurations that can be tested.
The simulators that support some of these features often require simulation of the entire network stack, even when a much smaller number of layers is sufficient.
Others do not support simulation of protocols that deviate too far from the IP stack -- which are those that we are most interested in testing and optimizing.

In contrast, the \dsnsfull{} supports arbitrary mobility and connectivity models, including fixed networks, LEO constellations, and interplanetary networking.
The links between nodes can be defined from a fixed list of edges or programmatically, based on distance, line of sight, angle of elevation, or planetary occlusion.
Its event-based architecture with optional minimum timesteps enables large spans of time to be fast-forwarded when messages travel a long distance alongside simulating large numbers of messages across shorter links within the same simulation scenario.
It is easily extensible through a simple Python interface, scalable to many thousands of nodes (demonstrated in Section~\ref{sec:evaluation-performance}), and simulates an abstracted network stack to improve efficiency and simplify development and iteration, while also supporting full-stack implementation of protocols if needed.

\section{Simulator Design}\label{sec:simulator-design}

\begin{figure}
    \centering\includegraphics[width=\linewidth]{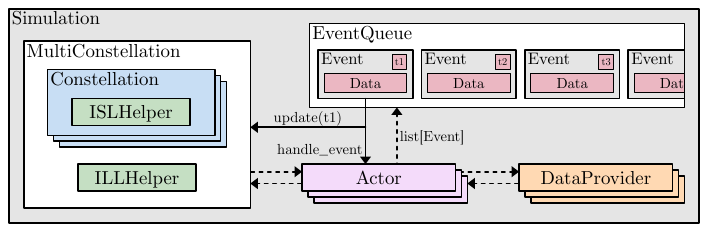}
    \caption{Overall structure of \dsns{} and its high-level operation.}
    \label{fig:dsns-architecture}
\end{figure}

In this section we describe the design of the \dsnsfull{}, demonstrating its underlying functionality, usage, and the ways in which it fulfills each of the requirements established in Section~\ref{sec:requirements}.

\subsection{Architecture}

Figure~\ref{fig:dsns-architecture} shows the overall architecture of \dsns{}.
The simulation is highly modular, underpinned by a simple event-based simulation: components can be switched out to alter behavior, or entirely new components can be added for additional functionality.
Mobility and connectivity are managed by the \texttt{MultiConstellation} class and its components, and all other functionality is handled by an event-based simulation and associated \texttt{Actors} and \texttt{DataProviders}.
Everything in the simulation, including message creation/routing/delivery, link changes, routing table updates, and scenario-specific changes, are stored in a priority queue as \texttt{Events} with attached timestamps.
These events are processed by \texttt{Actors}, which implement all complex functionality by matching and processing events -- for example, a single instance of the \texttt{MessageRoutingActor} handles message routing and delivery for all nodes in the simulation.

At each step of the simulation, the topmost event is removed from the queue and passed to all actors, which use pattern matching to decide whether to process the event, and optionally add further events to the queue.
They may also query data providers to gain additional information (e.g., routing tables), or the mobility model to get distances between nodes and the state of links.
This process repeats until the simulation terminates.
The modular implementation of complex functionality and protocols makes it easy to add new features without requiring a deep understanding of the rest of the simulation.

This architecture is highly efficient, as only the required components of the stack are simulated and simulation time can be advanced by large steps if needed. The latter is particularly useful in interplanetary settings as these often have long periods of little to no activity while a relay link is unavailable.
To further improve efficiency for high-traffic scenarios, we also provide an optional minimum time delta: with this setting, mobility and routing models will not be updated more than once within this timespan, reducing the amount of unnecessary updates, thus satisfying \ref{itm:dynamic-timesteps}.
The actor model also makes \dsns{} highly extensible (\ref{itm:extensible}), as we demonstrate through our implementation of LTP in Section~\ref{sec:protocol-design} -- we also explore further extension opportunities in Section~\ref{sec:extending-dsns}.

\subsection{Mobility}

Mobility and connectivity are handled by the \texttt{Multi\-Constellation} class, which simulates any number of constellations, the Inter-Satellite Links (ISLs) within constellations, and the Inter-Layer Links (ILLs) between constellations.
Constellations can be created from fixed points, Walker constellation parameters, or by importing Two Line Element (TLE) sets for full orbital simulation using the SGP4 propagator~\cite{valladoSGP42008}.
Each of these inherits from the \texttt{Constellation} class, defining the positions of satellites in the constellation or segment over time.
Each constellation also has an \texttt{OrbitalCenter}, defined as the parent around which satellites orbit, enabling complex movement and simulation of satellites around many different planets or other bodies, without loss of precision when communicating over short distances.
This satisfies \ref{itm:leo} and \ref{itm:ipn}.

Inter-satellite links can be fixed (for ground systems or Walker constellations) or dynamic (connecting to satellites within view), by using one of the pre-built \texttt{ISLHelper} classes or defining the links manually.
The \texttt{ILLHelper} works almost identically, but defines instead the links between different constellations or planets.
Each time the simulation time is updated, positions of satellites and connectivity of links are updated to reflect the new state -- satisfying \ref{itm:dynamic-connectivity}.

\subsection{Message Delivery}

Message routing and delivery in \dsns{} is modeled realistically by simulating propagation along each link in the network, for both point-to-point and broadcast messages.
This is handled for all nodes in the simulation by a \texttt{MessageRoutingActor} that, when a message is received, figures out the next hop in its path and forwards it on.
Propagation delays are modeled using speed-of-light distances between nodes, and the \texttt{LinkTransmissionActor} manages link bandwidth and transmission delays, queueing messages if the link is busy or buffering them if the link goes down.
All parameters including routing strategy, bandwidth, and error rate can be customized on a global or per-link basis.

Routing is handled by a \texttt{RoutingDataProvider}, which builds a connectivity graph for the network and computes optimal next-hop routing.
We provide routing systems for best-effort and store-and-forward delivery -- the latter looks ahead to future states of the network, enabling messages to be stored until the link becomes available.
We discuss in Section~\ref{sec:discussion} how future work can build upon this system to provide realistic implementations of current and proposed routing protocols.

\subsection{Protocol Design}\label{sec:protocol-design}

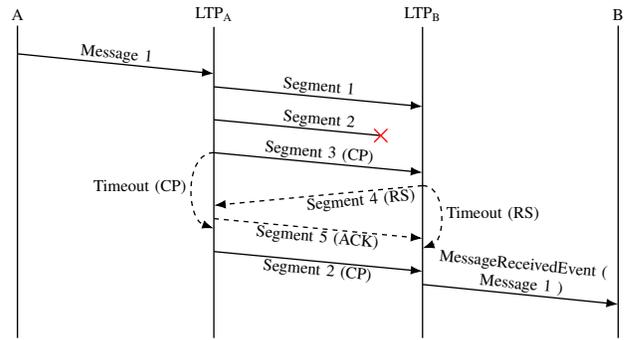
\begin{figure}
    \autobox{
    \begin{tikzpicture}[node distance=3.4cm,auto,>=stealth']
        \node[] (a) {A};
        \node[right = of a] (ltpa) {LTP\textsubscript{A}};
        \node[right = of ltpa] (ltpb) {LTP\textsubscript{B}};
        \node[right = of ltpb] (b) {B};
        \node[below of=a, node distance=7cm] (a_ground) {};
        \node[below of=ltpa, node distance=7cm] (ltpa_ground) {};
        \node[below of=ltpb, node distance=7cm] (ltpb_ground) {};
        \node[below of=b, node distance=7cm] (b_ground) {};
        \draw[thick] (a) -- (a_ground);
        \draw[thick] (ltpa) -- (ltpa_ground);
        \draw[thick] (ltpb) -- (ltpb_ground);
        \draw[thick] (b) -- (b_ground);
        
        \draw[thick,-{Latex[length=2.5mm]},sloped] ($(a)!0.12!(a_ground)$) -- node[above=-2.5pt,scale=1,midway]{\parbox{5cm}{\centering Message 1}} ($(ltpa)!0.18!(ltpa_ground)$);
        \draw[thick,-{Latex[length=2.5mm]},sloped] ($(ltpa)!0.22!(ltpa_ground)$) -- node[above=-2.5pt,scale=1,midway]{\parbox{5cm}{\centering Segment 1}} ($(ltpb)!0.28!(ltpb_ground)$);
        \coordinate (segment2_start) at ($(ltpa)!0.32!(ltpa_ground)$);
        \coordinate (segment2_end) at ($(ltpb)!0.38!(ltpb_ground)$);
        \coordinate (segment2_earlyend) at ($(segment2_start)!0.8!(segment2_end)$);
        \draw[thick,sloped] ($(segment2_start)$) -- node[above=-2.5pt,scale=1,pos=0.625]{\parbox{5cm}{\centering Segment 2}} ($(segment2_earlyend)$);
        \draw[red,thick] (segment2_earlyend)+(4pt,4pt) -- ++(-4pt,-4pt);
        \draw[red,thick] (segment2_earlyend)+(4pt,-4pt) -- ++(-4pt,4pt);
        \draw[thick,-{Latex[length=2.5mm]},sloped] ($(ltpa)!0.42!(ltpa_ground)$) -- node[above=-2.5pt,scale=1,midway]{\parbox{5cm}{\centering Segment 3 (CP)}} ($(ltpb)!0.48!(ltpb_ground)$);
        \draw[thick,dashed,-{Latex[length=2.5mm]}] ($(ltpa)!0.42!(ltpa_ground)$) to[out=180,in=160] node[pos=0.5,left]{Timeout (CP)} ($(ltpa)!0.65!(ltpa_ground)$);
        \draw[thick,dashed,-{Latex[length=2.5mm]}] ($(ltpb)!0.52!(ltpb_ground)$) to[out=0,in=20] node[pos=0.5,right]{Timeout (RS)} ($(ltpb)!0.71!(ltpb_ground)$);
        \draw[thick,dashed,-{Latex[length=2.5mm]},sloped] ($(ltpb)!0.52!(ltpb_ground)$) -- node[below=-2.5pt,scale=1,pos=0.3]{\parbox{5cm}{\centering Segment 4 (RS)}} ($(ltpa)!0.58!(ltpa_ground)$);
        \draw[thick,dashed,-{Latex[length=2.5mm]},sloped] ($(ltpa)!0.62!(ltpa_ground)$) -- node[below=-2.5pt,scale=1,midway]{\parbox{5cm}{\centering Segment 5 (ACK)}} ($(ltpb)!0.68!(ltpb_ground)$);
        \draw[thick,-{Latex[length=2.5mm]},sloped] ($(ltpa)!0.72!(ltpa_ground)$) -- node[below=-2.5pt,scale=1,midway]{\parbox{5cm}{\centering Segment 2 (CP)}} ($(ltpb)!0.78!(ltpb_ground)$);
        \draw[thick,-{Latex[length=2.5mm]},sloped] ($(ltpb)!0.82!(ltpb_ground)$) -- node[above=-2.5pt,scale=1,midway]{\parbox{5cm}{\centering MessageReceivedEvent (\\Message 1 )}} ($(b)!0.88!(b_ground)$);
    \end{tikzpicture}
    }
    \caption{Sequence diagram for LTP message retransmission. Segment 2 is lost in-transit and retransmitted to guarantee the message reaches its destination.}
    \label{fig:ltp-sequence-diagram}
\end{figure}

It is easy to extend \dsns{} to support new protocols (\ref{itm:extensible}); we demonstrate this by implementing the Licklider Transmission Protocol (LTP) for reliable per-hop message transmission, demonstrated in Figure~\ref{fig:ltp-sequence-diagram}~\cite{burleighLicklider2008,burleighSecurity2008}.
All functionality is contained within the \texttt{LTPActor} -- when enabled, this actor breaks the underlying message into green (unreliable) followed by red (reliable) \texttt{LTPDataSegments} and queues them for transmission.
This segmentation follows a configurable maximum segment size.
The last red data segment (if any) is marked as a checkpoint.
This aligns with the mandatory checkpoint requirements of the protocol, while support for optional discretionary checkpoints is left to future work.

Each segment is received through a \texttt{LTPSegment\- ReceivedEvent} and buffered until message reassembly is invoked. This in turn occurs when all expected red \texttt{LTPDataSegments} are received, or in the case of an all-green message, an end-of-block green \texttt{LTPDataSegment} is received.
At that point, the underlying message is reassembled and a \texttt{MessageReceivedEvent} is emitted.
Although standard LTP uses block offsets and lengths to compute missing byte ranges, our implementation abstracts this by having the checkpoint segment explicitly list the UIDs of sent \texttt{LTPDataSegments}, simplifying the receiver logic.
Upon receiving a checkpoint segment, the receiver responds with a \texttt{LTPReportSegment} listing the UIDs of the \texttt{LTPDataSegments} it received.
When the sender receives a \texttt{LTPReportSegment}, it sends a \texttt{LTPReportAcknowledgementSegment} and retransmits any missing \texttt{LTPDataSegments}.

Both the checkpoint and report segments have configurable timeouts, which trigger retransmissions if the corresponding responses are not received within the allotted time.
The actor can also be configured with a maximum number of retransmission attempts for each checkpoint or report segment.
If this limit is exceeded, the session is canceled and the message is dropped at the sender (\texttt{MessageDroppedEvent}) and/or has its reception aborted at the receiver (\texttt{MessageReceptionCanceledEvent}).

\subsection{Visualization}

\dsns{} also contains a visualization tool using the ``pyrender'' library, enabling a 3D view of any interplanetary network to provide a better understanding of how satellite network topologies evolve over time, and assist in the construction of new constellations or simulations.
This tool was used to create the visualization of an Earth-Mars network in Figure~\ref{fig:dsns-visualizer}.

\section{Reference Scenarios}\label{sec:reference-scenarios}

In this section we describe and implement a number of reference scenarios to demonstrate the capabilities and performance of \dsns{}, which can be used as a starting point for protocol development and optimization.
These scenarios define a network topology, bandwidth limitations, traffic models, and an error model, each of which can be mixed and matched, modified, or extended.

Each scenario is constructed from a simple Python script, examples of which are given in Appendix~\ref{app:example-code}.
Upon completion, they produce a log file detailing all events generated during the simulation (or aggregate statistics), from which further information can be extracted.
Metrics include message latency and hop count, bandwidth usage, and link saturation.
These can be used to assess the performance of different protocols or configurations, or to see how the characteristics of a given network topology change over time.

Finally, we also demonstrate the performance and scalability of \dsns{} itself (satisfying \ref{itm:scalable}), by running simulations with large numbers of nodes and high levels of traffic, and comparing to performance figures quoted for other simulators.

\subsection{CCSDS Reference Scenarios}

We start by implementing the network topologies specified in the DTN reference scenarios proposed by CCSDS~\cite{ccsdsReference2023}.\footnote{These scenarios are currently in draft form; our implementations in \dsns{} will be updated to reflect the final document.}
In these scenarios, the nodes, links, data rates, and traffic types are all well-defined, enabling consistent implementation for testing and development purposes.
The document specifies three scenarios:
\begin{itemize}
    \item \textbf{Earth Observation}: A payload control center and mission control center are each connected to two ground stations, which connect to an Earth observation satellite.
    \item \textbf{Lunar Communication}: A lunar base, rover, two relay satellites and a lunar gateway communicate with control centers and ground stations based on Earth.
    \item \textbf{Mars Communication}: Two rovers and three relay satellites are based on and around Mars, communicating with Earth-based control centers and ground stations.
\end{itemize}

\subsection{Custom Scenarios}

\begin{figure*}
    \centering
  \subfloat[Message delivery rate under each of the three message delivery systems.\label{fig:delivery-rates}]{%
       \includegraphics[width=0.325\linewidth]{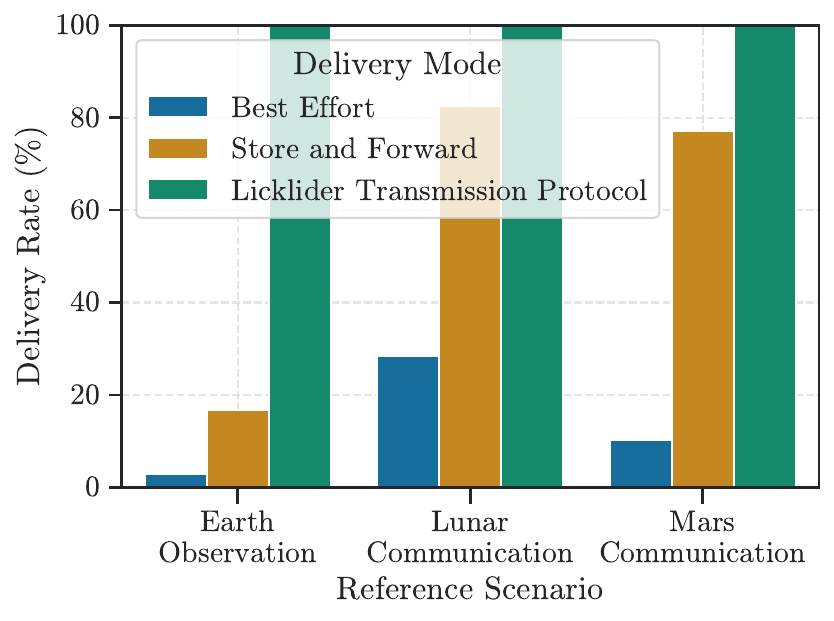}}
    \hfill
  \subfloat[Distribution of the latency of delivered messages, under the Earth Observation scenario, for each message delivery system.\label{fig:latencies-earth-observation}]{%
       \includegraphics[trim={0 -1.5em 0 0},width=0.325\linewidth]{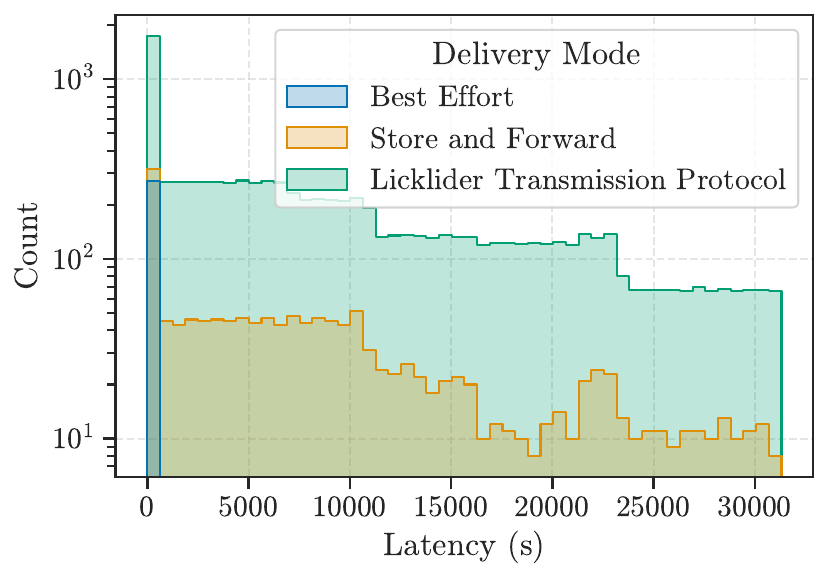}}
    \hfill
  \subfloat[Saturation of the three ground-to-space links over time, under the Earth Observation scenario with store-and-forward delivery.\label{fig:link-utilization-earth-observation}]{%
      \includegraphics[trim={0 -3.5em 0 0},width=0.325\linewidth]{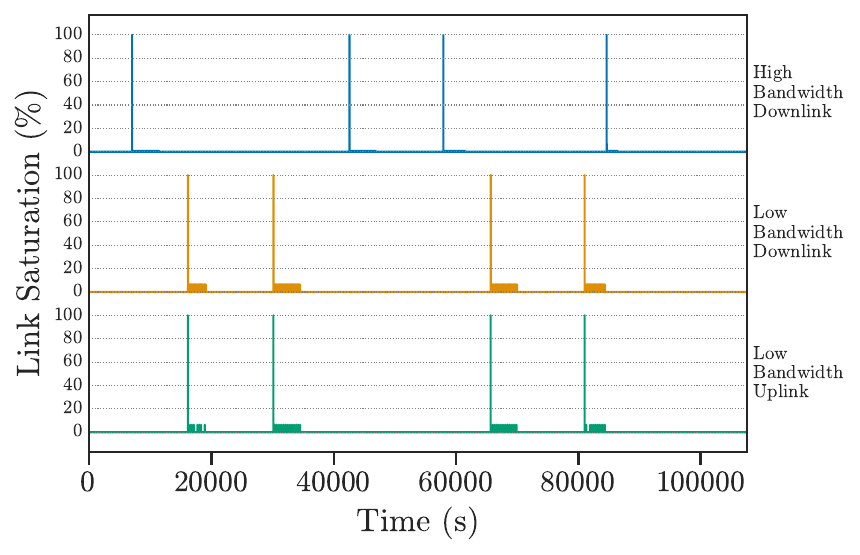}}
    \caption{A selection of results from the reference scenarios, demonstrating delivery rates, message latency, and link saturation over time.}
    \label{fig:reference-scenario-results}
\end{figure*}

Although useful for testing protocols against long- and short-distance communication, the above reference scenarios are very small in scale.
We therefore provide our own custom scenarios alongside these, focusing in particular on the scalability of \dsns{} applied to interplanetary networks -- the number of devices in each scenario can be scaled to suit the use case.
We implement the following scenarios:
\begin{itemize}
    \item \textbf{Walker Constellation}: A Walker constellation in LEO around Earth is connected to \num{12} ground stations and within itself via ISLs in a plus-grid topology. By default we use \num{66}~satellites matching the Iridium constellation.
    \item \textbf{CubeSat Constellation}: Using data from CelesTrak~\cite{celestrak1985} and the SGP4 orbital propagator~\cite{valladoSGP42008}, we construct a federated network composed of the \num{98} CubeSats currently in orbit, connected to \num{12} ground stations, and to each other via opportunistic ISLs with a range of \qty{2500}{\kilo\metre}.\footnote{We use data provided on 2025-06-27 and propagated from 2025-06-27T00:00:00Z. Up-to-date TLEs can be substituted in if needed.}
    \item \textbf{Lunar-Mars Communication}: The ``Walker Constellation'' scenario is augmented with Lunar connectivity, with a Walker constellation of \num{8} satellites connected to \num{12} Lunar ground stations, and a single relay satellite connecting to three ground systems on Earth. A network around Mars is also added, with \num{66} satellites and \num{12} ground stations, and its own relay satellite.
\end{itemize}

For these scenarios, we assume a uniform data rate of \qty{25}{\mega\bit/\second} by default, matching the reported bandwidth of Iridium's ISLs~\cite{yangInterSatellite2025} -- as with the CCSDS scenarios, this can be configured on a global or per-link basis to match the parameters of a particular mission.

To ensure the scenarios are as versatile as possible, the traffic generation strategy can also be configured.
We provide the following options:
\begin{itemize}
    \item \textbf{Point-to-point Communication:} Pairs of nodes send data between each other at a constant rate for the full duration of the simulation. The specific pairs of nodes, rate of communication, and size of generated messages are all configurable. By default, 10 pairs of nodes send a \qty{1}{\mega\byte} block of data every second.
    \item \textbf{Random Communication}: Nodes randomly generate messages to send between each other. The message rate, source, destination, and size are all drawn from user-configurable distributions -- by default, we support constant, uniform, Gaussian, and Pareto distributions.
\end{itemize}

\subsection{Message Delivery}

When running any of the above reference scenarios, we must also choose the message delivery system which the simulation uses to route and deliver messages.
We have implemented three message delivery systems:
\begin{itemize}
    \item \textbf{Best-effort delivery:} Following the IP paradigm, messages are routed according to instantaneous route availability, and are dropped if no route is available.
    \item \textbf{Store-and-forward delivery:} Using a BP-style strategy, the routing system looks forward to the state of the network in the future to ensure messages arrive at their destination, storing messages if needed while waiting for a link to become available.
    \item \textbf{Store-and-forward with LTP:} The same store-and-forward strategy is used, but with Licklider Transmission Protocol implemented at the data link layer to ensure reliable data transport and realistic transmission queues.
\end{itemize}
Each of these can be swapped out with a single option.
We focus on demonstrating store-and-forward (with and without LTP) in our results, as best-effort delivery is not well-suited to networking in space, where links are often unavailable.

\subsection{Error Model}

Finally, we provide an error model with a configurable message loss rate.
By default, this is set to \qty{5}{\percent} across all links, but this can be adjusted globally or on a per-link basis.
These message errors usually result in failed delivery, unless reliable data transport is used -- in our results, we demonstrate how the introduction of LTP prevents message loss.

\section{Evaluation}\label{sec:evaluation}

In this section we run a number of reference scenarios to demonstrate the capability of \dsns{} in a wide range of contexts, and showcase its performance and scalability to large-scale constellations.

\subsection{Reference Scenarios}

By running simulations using the reference scenarios outlined above, we can demonstrate the effectiveness of \dsns{} in a wide range of contexts.
For this section we focus on the CCSDS reference scenarios; full results are  in Appendix~\ref{app:extended-results}.

\subsubsection{Delivery Rate and Latency}

We look first at the delivery rate of messages under each message delivery system; these results are summarized for each of the CCSDS scenarios in Figure~\ref{fig:delivery-rates}.
We can see that, as expected, delivery rate is much lower under best-effort delivery, since any messages that are generated when a link is not immediately available are discarded.
Delivery rates under store-and-forward delivery are better, but messages are still lost due to the \qty{5}{\percent} error rate.
However, using LTP successfully reduces the number of failed deliveries to \qty{0}{\percent}, since any lost data is retransmitted.

We gain further insights looking at the distribution of the latency of delivered messages, shown in Figure~\ref{fig:latencies-earth-observation}.
Latency under best-effort delivery is very low, since the only delivered messages are those for which no buffering is required.
LTP results in higher latencies than store-and-forward, since more messages are delivered instead of dropped, and additional overhead is incurred by retransmissions and acknowledgments.

\subsubsection{Link Saturation}

Next, we look at how the saturation of links changes over time, shown in Figure~\ref{fig:link-utilization-earth-observation} for the Earth Observation scenario under store-and-forward delivery, looking in particular at the ground-to-space links.
As expected, whenever a link comes online, we see a spike in saturation as all the buffered messages are sent, followed by the newly generated messages, resulting in a more consistent amount of low saturation.
In large-scale networks, this analysis is useful for locating the most saturated links in a network and tracking down its root cause.

\subsection{Performance}\label{sec:evaluation-performance}

\begin{figure}
    \centering\includegraphics[width=.8\linewidth]{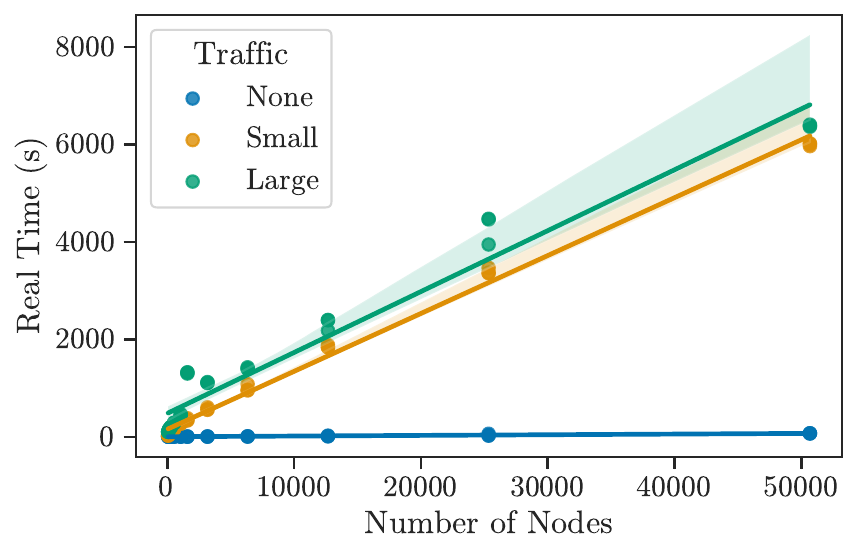}
    \caption{Time taken to run the Walker constellation scenario in \dsns{} as the number of nodes in the constellation increases.}
    \label{fig:performance-time}
\end{figure}

Finally, we use these scenarios to demonstrate the performance and scalability of \dsns{}~(\ref{itm:scalable}), and compare to other network simulators.
This analysis can be repeated for any network topology and traffic configuration; for the sake of simplicity we choose to focus on the Walker constellation with point-to-point communication, running for \num{100}~minutes of simulation time.
In order to get the best understanding of the overhead imposed by \dsns{} itself, we configure the simulator to use best-effort delivery with no lookahead in routing -- the topology of the Walker constellation does not change significantly, so it is not necessary to use lookahead.

We start with a simulation of \num{66}~nodes, gradually increasing the number of nodes to \num{50688}.
For each constellation, we run simulations with three different traffic configurations: no communication between nodes, low traffic communication (\qty{10}{\mega\byte} every \num{11.5}~seconds, matching the EOS scenario), and high traffic communication (\mathqty{100\times10}{\mega\byte} every \num{11.5}~seconds).
All experiments were executed on a single CPU core, using an Intel Xeon E5-2660 processor running at \qty{3.3}{\giga\hertz}.
No simulation used more than \qty{2.1}{\giga\byte} of RAM -- further details are in Appendix~\ref{app:extended-results}.

The time taken to run each simulation is summarized in Figure~\ref{fig:performance-time}.
We see that even on the larger traffic configuration, on networks whose scale far exceeds any existing satellite network, the simulation still runs faster than real time.
On smaller networks, or with smaller volumes of traffic, simulations are even faster, with many running tens or hundreds of times faster than real time.
This exceeds the performance of even the best-performing simulators in related work: ``Stardust'' takes just over \num{6000}~seconds to simulate \num{20600}~satellites~\cite{pusztaiStardust2025}, and does not perform full traffic simulation, and ``StarryNet'' takes over \num{3000}~seconds to simulate \num{1000}~satellites under the same conditions~\cite{pusztaiStardust2025}.
``Hypatia'' has a slowdown rate of between \num{2} and \num{50000} on a network of just over \num{1000} satellites~\cite{kassingExploring2020} depending on traffic volume, compared to \dsns{}' slowdown rate of less than \num{1} under all tested configurations.

Furthermore, since \dsns{} is single-threaded, many simulations can be executed at once on a single machine with no impact on performance for each simulation.
This is highly useful in practice, enabling multiple tests under different network topologies or protocol configurations to be executed at the same time and compared afterwards.

\section{Extensibility of \dsns{}}\label{sec:extending-dsns}

\begin{figure}
    \centering\includegraphics[width=\linewidth]{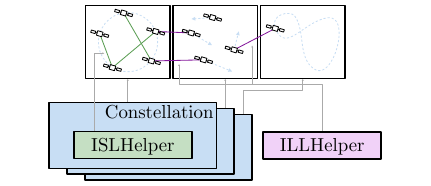}
    \caption{\dsns{} can be extended to support new mobility and connectivity models by building new instances of the \texttt{Constellation}, \texttt{ISLHelper}, and \texttt{ILLHelper} classes.}
    \label{fig:extending-dsns}
\end{figure}

One major use case for \dsns{} lies in its extensibility (\ref{itm:extensible}) -- each of the reference scenarios can be modified, extended, or rewritten to support the needs of a particular experimental goal.
By using the same reference scenarios and metrics across all experiments, we can be sure that new protocols or network topologies provide concrete improvements, rather than merely performing better as an artifact of the experimental setup.

Due to its modular design, adding new functionality to \dsns{} is a highly straightforward process.
We demonstrate this by outlining how a user might add support for a new network layer above or below those already simulated, a new routing strategy, and new constellations within the mobility and connectivity model.

Since the behavior of each component is defined by Python functions, rather than a domain-specific language, customization options are almost unlimited: as long as the desired functionality can be expressed in code, it can be supported in \dsns{}.
This enables a choice of different levels of abstraction based on need: for the majority of the examples expressed in this paper we have focused on abstracted implementations of protocols, allowing us to test functionality with minimal computational impact.
However, we have also seen through our implementation of LTP that protocols can be implemented with greater fidelity.
If required, this can even be taken down to the bit level for a given protocol -- and importantly, this can be done without requiring the rest of the stack to be simulated with such high precision, unless specifically required.

\subsection{Mobility and Connectivity}

Firstly, we consider extensions to the mobility and connectivity model used by \dsns{}, illustrated in Figure~\ref{fig:extending-dsns}.
These are the easiest parts of the simulator to modify, as they are unaffected by the rest of the simulation.
To add a new type of orbit, a user simply extends the \texttt{Constellation} class to provide the desired functionality, defining \texttt{update\_positions} to correctly set their positions at each timestep.
These can be fixed, follow a given curve or orbital propagation model, move randomly, or follow any other pattern that can be expressed in code.
Similarly, \texttt{ISLHelper} can be extended to provide new connectivity models, with \texttt{get\_isls} defining how satellites in a constellation are connected at any given point in time -- by modifying this function, any configuration of links can be supported, from fixed topologies to more esoteric options.
Finally, \texttt{ILLHelper} (and corresponding function \texttt{get\_ills}) defines the connections between layers and planetary segments, and can be modified in the same way to enable arbitrary inter-layer and inter-segment connectivity.

Once these have been defined, the simulator handles all low-level functionality, solving for routes and generating link up/down events as links become available and unavailable.

\subsection{Routing Strategies}

Implementing a new routing strategy in \dsns{} is straightforward at its core, requiring an extension to the \texttt{RoutingDataProvider} class with the new strategy.
However, this obscures some underlying complexity -- the existing routing strategies rely upon global knowledge of the network state, and do not require message passing between nodes.

To implement a routing protocol that involves an exchange of information between nodes, additional functionality must be integrated into the data provider.
This could be build on top of the existing message delivery system, using manually routed messages to pass information between adjacent nodes.

The message delivery system can also be extended to support new functionality; for instance, if the data provider returns more than one next hop, it could be configured to deliver messages along multiple paths to increase the chance of successful delivery.
This flexibility enables the implementation of a wide range of routing protocols and strategies.

\subsection{Higher Layer Protocols}

It is easy to add new protocols at higher levels by extending the \texttt{BaseMessage} to add additional fields, encapsulating the original message as a field within the new message.
However, developers need to take care that layers are appropriately ordered by ensuring each layer correctly pattern matches on messages from the layer above it, and that a \texttt{MessageCreatedEvent} is only generated for the lowest layer event.
For more complex simulations with multiple layers, a layer management actor might be used to make sure layers are correctly ordered and handle events between them.

\subsection{Physical Layer}

Alongside higher layer functionality, \dsns{} also supports implementing additional features at lower layers.
This might involve adding additional parameters such as message size, priority, or physical layer modeling for more realistic message loss.
To modify the message structure, \texttt{BaseMessage} can be extended once again to add the new fields.
Following this, the \texttt{MessageRoutingActor} can be extended to modify physical layer delivery mechanisms -- for example, the \texttt{handle\_message\_sent\_event} function can be modified to support bandwidth limiting or message loss modeling.

\subsection{Security}

Finally, we discuss the ways in which \dsns{} can be used to enable security research in this area.
In the source code for the simulator we provide mechanisms for simulating the following attacks:
\begin{itemize}
    \item \textbf{Global message loss:} We provide configurable randomized message loss across the whole network through the \texttt{LossConfig} class, enabling the testing of reliable delivery and rerouting mechanisms under realistic loss conditions.
    \item \textbf{Targeted message loss:} The \texttt{LossConfig} also supports per-link loss rates, enabling simulation of more targeted attacks and ensuring redelivery and rerouting still works under direct attack.
    \item \textbf{Link flooding attacks:} Finally, we provide tools to simulate targeted flooding attacks, denying service by exhausting the available bandwidth on a link. This can be accessed via the \texttt{TrafficFloodActor}, and can be configured to start and end at a specific time, target a single link or a group of links, and use different amounts of bandwidth.
\end{itemize}

Furthermore, the simplicity and extensibility of \dsns{} makes it easy to build further attack mechanisms, making it particularly useful to assess the behavior of different protocols and strategies under a wide range of attacks.

\section{Discussion}\label{sec:discussion}

In previous sections we have demonstrated that \dsns{} is efficient, scalable, supports any required network topology, and can be easily extended to support new protocols.
The combination of these features makes it immensely useful for future research, development, and testing.
Standards bodies like the CCSDS can use the simulator to run their DTN communication reference scenarios, with fast iteration on parameters for upcoming protocols and recommended standards.
They will also be able to run simulations at a much larger scale than these small examples, using configurations built upon the custom scenarios explored in this paper to test those same protocols on significantly larger networks without sacrificing performance.
In doing so, we can ensure standards are well-informed by simulations which match real-world configurations.

Alongside these standards bodies, satellite developers and operators will also benefit through making data-driven decisions prior to deployment.
Alongside optimizing their planned network topology, they will also be able to test protocols to ensure their configuration provides optimal performance.

Although the performance of \dsns{} is already sufficient for the vast majority of cases, there is always scope for further improvements in this area.
For example, the impact of message routing could be reduced by precomputing routes on paths that are known to be fixed, and components of the mobility and connectivity model could be written in a more performant language such as Rust or C++.
Thanks to the modularity of \dsns{} combined with its thorough documentation, it will be possible to replace components with higher-performance variants without impacting usage or results.

Future work might also consider implementing the JSON/CSV specifications used by the CCSDS reference scenarios, enabling them to be implemented and tweaked directly using the same format as the ESA simulation platform.
By running the same simulations on multiple different platforms, we can gain the best of both worlds -- the SpaceSecLab platform provides better support for full-stack network simulation, whereas \dsns{} supports larger constellations and faster iteration on protocol design, so both can be used together for effective testing across the board.
However, implementing this specification is not trivial, as \dsns{} makes use of full orbital simulation to calculate rendezvous times, whereas the CCSDS specification uses a list of fixed connection times.
Parity between the two can be ensured by either using \dsns{} to generate the rendezvous times for the CCSDS scenarios, or by modifying the simulation configuration for the reference scenarios to instead take a fixed list of connection times.

\section{Conclusion}\label{sec:conclusion}

In this paper we have presented the \dsnsfull{} (\dsns{}), and demonstrated how it can be used to enhance the research and development necessary for upcoming LEO megaconstellations and interplanetary networks.
Through its modular architecture, event-based simulation, and a simple and flexible interface, \dsns{} offers improved scalability, configurability, and extensibility compared to existing tools.
We have demonstrated its capabilities through the implementation of existing protocols and reference scenarios, and shown that its performance exceeds existing state-of-the-art tools.

As satellite networks continue to expand in scale and capability, \dsns{} is well-positioned to enable future innovation in this critical area, bringing us closer to the ultimate goal of a unified interplanetary internet.

\section*{Acknowledgments}
We would like to thank armasuisse Science + Technology for their support during this work.
Joshua was supported by the Engineering and Physical Sciences Research Council (EPSRC).
Sebastian and Simon were supported by the Royal Academy of Engineering and the Office of the Chief Science Adviser for National Security under the UK Intelligence Community Postdoctoral Research Fellowships programme.

\clearpage
\printbibliography

\clearpage
\appendices
\section{Example Code}\label{app:example-code}

In this appendix, we provide code snippets illustrating the ease with which scenarios can be constructed using \dsns{}, the reference scenarios described in Section~\ref{sec:reference-scenarios} as a base.

The following builds and runs the Earth Observation scenario:
\begin{minted}[frame=lines, framesep=2mm, baselinestretch=1.2, fontsize=\footnotesize, linenos]{python}
constellation = EarthObservationMultiConstellation()

transmission_actor = \
EarthObservationTransmissionActor(
    constellation=constellation
)

traffic_actor = EarthObservationTrafficActor(
    constellation=constellation,
    update_interval=300,
)

routing_data_provider = LookaheadRoutingDataProvider(
    resolution=60,
    num_steps=600,
)
routing_actor = MessageRoutingActor(
    routing_data_provider,
    store_and_forward=True,
    model_bandwidth=True,
    #loss_config=LossConfig(
    #    seed=0,
    #    default_loss_probability=0.05
    #),
    #reliable_transfer_config=LTPConfig(),
)

simulation = Simulation(
    constellation,
    actors=[
        transmission_actor,
        traffic_actor,
        routing_actor,
    ],
    data_providers=[routing_data_provider],
    timestep=0.01,
)

simulation.initialize(time=0)
simulation.run(3600*24, progress=False)
\end{minted}

To modify this scenario to use LTP, we simply uncomment the \texttt{LTPConfig()}, alongside the following LTP actor:
\begin{minted}[frame=lines, framesep=2mm, baselinestretch=1.2, fontsize=\footnotesize, linenos]{python}
retransmission_config = RetransmissionConfig()
ltp_actor = LTPActor(
    config=retransmission_config,
    model_bandwidth=True,
)
\end{minted}

Similarly, message loss can be added by uncommenting the \texttt{LossConfig()}.

We provide the helper scripts \texttt{ccsds\_reference.py} and \texttt{custom\_reference.py} to build and run all of the scenarios described in this paper and collect logs, with a range of options to customize the simulation.
These can be used as they are, or as a starting point for modification.

\section{Extended Results}\label{app:extended-results}

In this appendix we expand upon the simulation results in Section~\ref{sec:evaluation}, providing additional statistics and insights.

\subsection{Reference Scenarios}

\begin{table*}
    \caption{Aggregated results for each of the tested reference scenarios.}
    \label{tab:results-summary}
    \autobox{
    \begin{tabular}{llS[table-format=1.1]S[table-format=3.2]S[table-format=2.2]S[table-format=5.2]S[table-format=2.2]S[table-format=3.2]S[table-format=2.2]}
    \toprule
    \multicolumn{3}{c}{Configuration} \\
    \cmidrule(lr){1-3}
    Scenario & Delivery Mode & {Loss (\%)} & {Delivered (\%)} & {Dropped (\%)} & {Mean Latency (s)} & {Mean Hops} & {\begin{tabular}[b]{@{}c@{}}Mean Link\\Utilization (\%)\end{tabular}} & {\begin{tabular}[b]{@{}c@{}}Max Link\\Utilization (\%)\end{tabular}} \\
    \midrule
    Earth Observation & Best Effort & 5.0 & 2.93 & 83.35 & 0.10 & 2.00 & 1.48 & 6.40 \\
    Earth Observation & Store and Forward & 5.0 & 16.76 & 0.00 & 8986.45 & 2.00 & 7.79 & 100.00 \\
    Earth Observation & Licklider Transmission Protocol & 5.0 & 100.00 & 0.00 & 9704.86 & 2.00 & 9.29 & 100.00 \\
    Lunar Communication & Best Effort & 5.0 & 28.51 & 60.03 & 1.31 & 3.35 & 0.99 & 17.07 \\
    Lunar Communication & Store and Forward & 5.0 & 82.62 & 0.00 & 4707.89 & 3.70 & 3.22 & 100.00 \\
    Lunar Communication & Licklider Transmission Protocol & 5.0 & 100.00 & 0.00 & 4733.25 & 3.71 & 1.83 & 100.00 \\
    Mars Communication & Best Effort & 5.0 & 10.34 & 86.89 & 1259.66 & 4.65 & 1.42 & 51.20 \\
    Mars Communication & Store and Forward & 5.0 & 77.16 & 0.00 & 14732.93 & 5.05 & 13.06 & 100.00 \\
    Mars Communication & Licklider Transmission Protocol & 5.0 & 100.00 & 0.20 & 14517.87 & 4.97 & 9.65 & 100.00 \\[.4em]
    Walker Constellation & Best Effort & 0.5 & 67.58 & 0.35 & 2.01 & 4.73 & 28.47 & 67.59 \\
    Walker Constellation & Store and Forward & 0.5 & 68.52 & 0.00 & 4.67 & 4.73 & 29.20 & 100.00 \\
    Walker Constellation & Licklider Transmission Protocol & 0.5 & 100.00 & 0.00 & 3.42 & 4.96 & 15.25 & 100.00 \\
    CubeSat Constellation & Best Effort & 0.5 & 32.63 & 42.13 & 2.64 & 5.70 & 28.23 & 100.00 \\
    CubeSat Constellation & Store and Forward & 0.5 & 49.43 & 1.20 & 53.38 & 7.36 & 35.04 & 100.00 \\
    CubeSat Constellation & Licklider Transmission Protocol & 0.5 & 98.80 & 1.20 & 78.25 & 10.02 & 21.03 & 100.00 \\
    Lunar-Mars Communication & Best Effort & 0.5 & 59.72 & 8.65 & 102.52 & 5.12 & 27.64 & 67.59 \\
    Lunar-Mars Communication & Store and Forward & 0.5 & 64.40 & 0.00 & 239.36 & 5.44 & 29.14 & 100.00 \\
    Lunar-Mars Communication & Licklider Transmission Protocol & 0.5 & 100.00 & 0.00 & 421.20 & 6.04 & 16.17 & 100.00 \\
    \bottomrule
    \end{tabular}
    }
\end{table*}

\begin{figure*}
    \centering
  \subfloat[Message delivery rate under each of the three ``custom'' message delivery systems.\label{fig:delivery-rates-custom}]{%
       \includegraphics[width=0.325\linewidth]{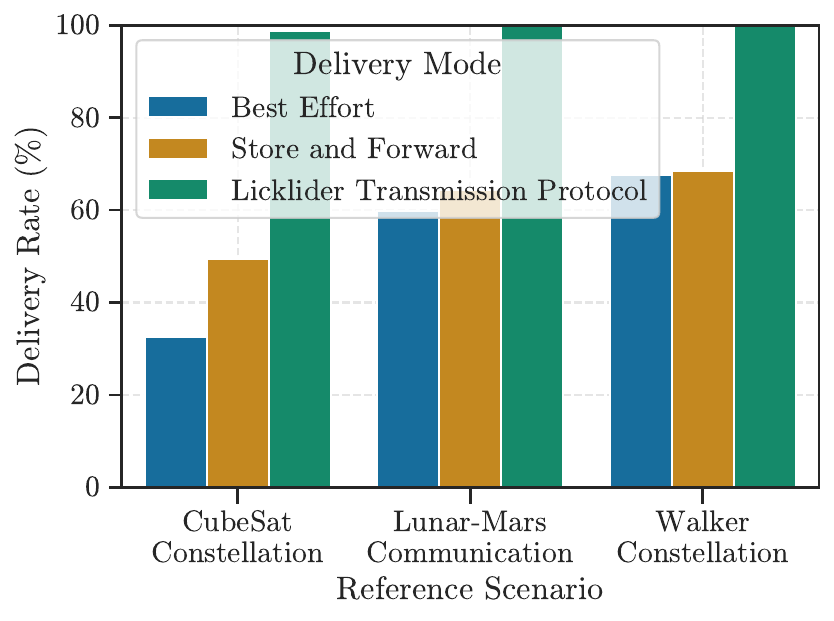}}
    \hfill
  \subfloat[Distribution of the latency of delivered messages, under the Walker Constellation scenario, for each message delivery system.\label{fig:latencies-walker}]{%
        \includegraphics[trim={0 -1.5em 0 0},width=0.325\linewidth]{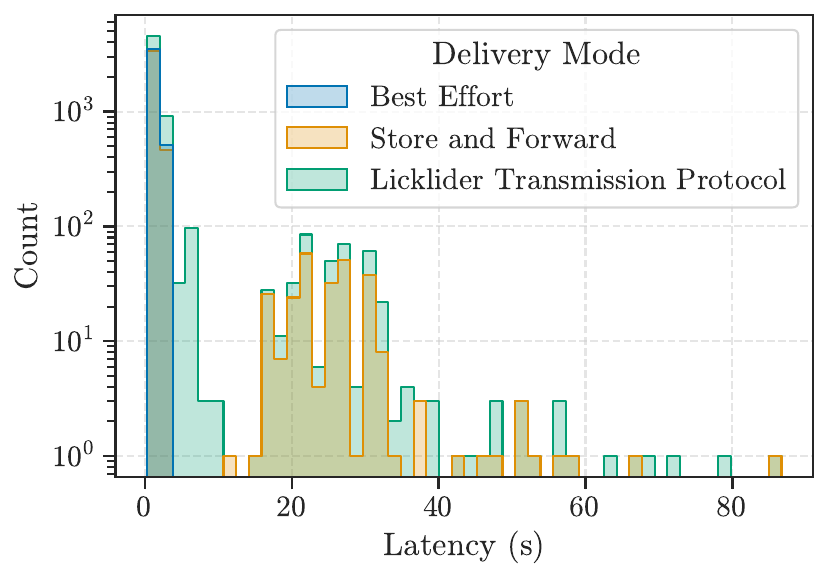}}
    \hfill
  \subfloat[Saturation of the Earth-Mars links over time, under the Lunar-Mars Communication scenario with LTP delivery.\label{fig:link-utilization-lunar-mars}]{%
        \includegraphics[trim={0 -3.5em 0 0},width=0.325\linewidth]{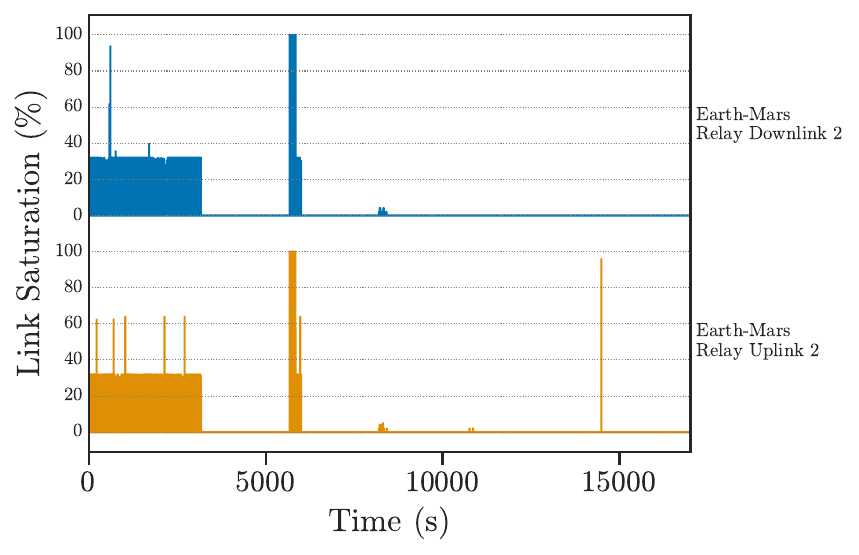}}
    \caption{A selection of results from the reference scenarios, demonstrating delivery rates, message latency, and link saturation over time.}
    \label{fig:reference-scenario-results-custom}
\end{figure*}

In addition to the CCSDS reference scenarios, we also ran simulations under the custom reference scenarios provided in this paper.\footnote{We adjust the loss rate for these scenarios to \qty{0.5}{\percent}.}
Results for all the scenarios are summarized in Table~\ref{tab:results-summary},\footnote{Any messages that are not delivered and not dropped (e.g., due to not being able to find a route for the message) are lost due to the scenario's error model.} and a selection of the results are given in Figure~\ref{fig:reference-scenario-results-custom}.
We can see that when LTP is not used, the delivery rates are lower, even with the lowered error rate -- this is due to the fact that messages must traverse many more links to reach their destination, causing the error rate to compound over each consecutive link.
However, the inclusion of LTP counteracts this and ensures almost all messages are delivered successfully.
We also see that delivery rates for best-effort and store-and-forward delivery are closer to one another, particularly in the Walker constellation scenario, since connectivity is largely fixed except for ground-to-space links, so the majority of losses are due to the error model.

\subsection{Performance}

\begin{figure}
    \centering\includegraphics[width=\linewidth]{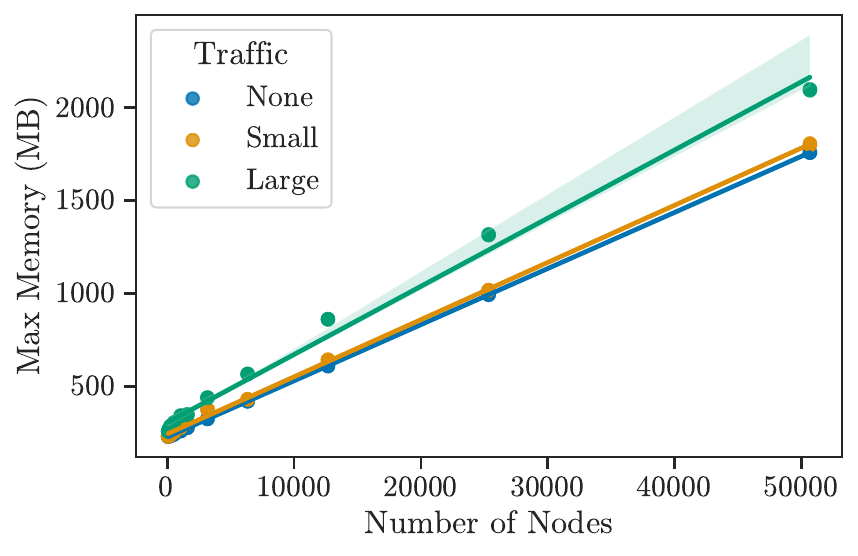}
    \caption{Maximum memory used by \dsns{} while running the Walker constellation scenario as the number of nodes in the constellation increases.}
    \label{fig:performance-memory}
\end{figure}

Figure~\ref{fig:performance-memory} shows the memory used by \dsns{} for each of the Walker simulations.
We can see that it increases linearly with the number of nodes in the simulation, due to the additional data associated with each node (position, routing tables, etc.) and increases slightly with the volume of traffic in the simulation.
It is possible that memory usage could be reduced through the use of profilers and efficiency improvements, but it is already low enough for the vast majority of use cases -- even on the largest simulations, usage did not exceed \qty{2.1}{\giga\byte}.

\end{document}